\documentclass[aps,showpacs,nofootinbib,twocolumn,preprintnumbers,nobalancelastpage,superscriptaddress]{revtex4}

\usepackage[english]{babel}
\usepackage{amsmath,amssymb}
\usepackage[dvips]{graphicx}
\usepackage[sort&compress]{natbib}
\usepackage{amsfonts}
\usepackage{bbm}
\usepackage{color}

\DeclareMathAlphabet{\mathsc}{OT1}{cmr}{m}{sc}

\allowdisplaybreaks

\def\321{SU(3) $\otimes$ SU(2) $\otimes$ U(1)}

\def\lsim{\raise0.3ex\hbox{$\;<$\kern-0.75em\raise-1.1ex\hbox{$\sim\;$}}}
\def\gsim{\raise0.3ex\hbox{$\;>$\kern-0.75em\raise-1.1ex\hbox{$\sim\;$}}}

\newcommand{\flux}[2][]{\ensuremath{\ifthenelse{\equal{#1}{}}{}{^{#1}\!}\mathit{#2}}}

\newcommand{\sla}[1]{/\!\!\!\!#1}

\begin{document}
\preprint{YITP-SB-09-22}

\title{Signals for New Spin-1 Resonances in Electroweak 
Gauge Boson Pair Production at the LHC}
\author{A.\ Alves}
\email{aalves@fma.if.usp.br}
\affiliation{Instituto de F\'{\i}sica Te\'orica,
             Universidade Estadual Paulista, S\~ao Paulo -- SP, Brazil.}

\author{O.\ J.\ P.\ \'Eboli}
\email{eboli@fma.if.usp.br}
\affiliation{Instituto de F\'{\i}sica,
             Universidade de S\~ao Paulo, S\~ao Paulo -- SP, Brazil.}

\author{D.\ Gon\c{c}alves Netto}
\email{dorival@fma.if.usp.br}
\affiliation{Instituto de F\'{\i}sica,
             Universidade de S\~ao Paulo, S\~ao Paulo -- SP, Brazil.}

\author{M.\ C.\ Gonzalez--Garcia}
\email{concha@insti.physics.sunysb.edu}
\affiliation{%
  C.N.~Yang Institute for Theoretical Physics,
  SUNY at Stony Brook, Stony Brook, NY 11794-3840, USA
  \\
  Instituci\'o Catalana de Recerca i Estudis Avan\c{c}ats (ICREA),
  Departament d'Estructura i Constituents de la Mat\`eria, Universitat
  de Barcelona, 647 Diagonal, E-08028 Barcelona, Spain}

\author{J.\ K.\ Mizukoshi}
\email{mizuka@ufabc.edu.br}
\affiliation{Centro de Ci\^encias Naturais e Humanas, Universidade Federal do 
ABC, Santo Andr\'e -- SP, Brazil.}

\begin{abstract}
  The mechanism of electroweak symmetry breaking (EWSB) will be
  directly scrutinized soon at the CERN Large Hadron Collider
  (LHC). We analyze the LHC potential to look for new vector bosons
  associated with the EWSB sector, presenting a possible model independent 
  approach to search for these new spin--1 resonances. We show that the 
  analyses of the processes $pp \to \ell^+ \ell^{\prime -} \, \sla{E}_T, 
  \; \ell^\pm jj \, \sla{E}_T, \; \ell^{\prime \pm} \ell^+ \ell^-  \,\sla{E}_T, \; 
  \ell^\pm jj \, \sla{E}_T$, and $\ell^+ \ell^- jj$ (with $\ell, \ell^\prime = e$ 
  or $\mu$ and $j =$ jet) have a large reach at the LHC and can lead to the 
  discovery or exclusion of many EWSB scenarios such as Higgsless models.
\end{abstract}

\pacs{ 95.30.Cq} 

\maketitle


\section{Introduction}

The CERN Large Hadron Collider is about to start probing directly the
TeV scale with the study of the electroweak symmetry breaking
mechanism being at center stage. In order to respect unitarity in the
weak gauge boson scattering $W^+_L W^-_L \to W^+_L W^-_L$ there must
be a contribution of the EWSB at the TeV scale~\cite{Lee:1977eg} ,
well within the LHC reach.  In the SM the growth of the scattering
amplitude of this process is cutoff by the presence of the Higgs
boson, however, the simplest one doublet Higgs sector is just one of a
myriad of possibilities for the symmetry breaking sector of the
electroweak interactions. \medskip

One appealing possibility is that the electroweak symmetry breaking is
caused by a new strongly interacting sector~\cite{TC}.  Although
simple scaled up QCD models are ruled out by precision measurements
and flavor changing neutral current constraints, it is possible to
build viable models of dynamical electroweak symmetry
breaking~\cite{NTC}. A common feature of many of these models is the
appearance of new spin--1 states that unitarize the weak gauge boson
scattering. \medskip

Recently scenarios presenting extra dimensions have received a lot of
attention, allowing the construction of Higgsless models where
unitarity restoration takes place through the exchange of an infinite
tower of spin--1 Kaluza-Klein (KK) excitations of the known
electroweak gauge bosons~\cite{Csaki:2003dt}.  In this class of
models~\cite{hless} unitarity constraints give rise to sum rules
depending on the couplings and masses of the KK excitations that imply
that the first KK excitation should be observable at the LHC, while
higher KK modes are probably beyond the LHC
reach~\cite{Birkedal:2004au}. \medskip

A common feature of many EWSB scenarios, as the ones above mentioned,
is the existence of new vector resonances, $Z^\prime$ and $W^\prime$, that couple
to $W^+W^-$ and $W^\pm Z$ pairs, respectively. But, generically, their
properties, such as mass, width, and couplings to SM fermions, are
model dependent.  In this respect, the most {\sl model independent}
channel for detection of such spin-1 resonances would be the
observation of their contribution to SM gauge boson fusion (WBF) which
only involves their couplings to electroweak gauge bosons. At the LHC
this would correspond to the $Z^\prime$ exchange contribution to $W^+W^-
\rightarrow W^+W^-$, such as for example in the process $p
p\rightarrow j j \ell^+ \ell^{\prime -} \,\sla{E_T}$.  Unfortunately, that
signal is unobservable at LHC even with increased luminosity
\cite{Birkedal:2004au,Alves:2008up}. On the other hand, $W^\prime$ can be
observed in the WBF $W^\pm Z \to W^\pm Z$ elastic scattering
~\cite{Birkedal:2004au,Alves:2008up}.

The new spin-1 states can also be directly produced in $pp$ collisions
via their  couplings to light quarks. In this case, in order to determine 
that such new vector bosons are indeed associated with EWSB one can study   
processes in which the new spin-1 decays into electroweak gauge boson
pairs,
{\em i.e.}  
\begin{eqnarray}
&&\!\!\!\!\!\!\!\!\!\!\!\!
p p \to Z^\prime \to W^+ W^- \to \ell^+ \ell^{\prime -} \, \sla{E}_T \hbox{ and } 
\ell^\pm jj \, \sla{E}_T
\label{ppww}
\\
\nonumber
\\
&&\!\!\!\!\!\! \!\!\!\!\!\! 
p p \to W^\prime \to W^\pm Z \to \ell^{\prime \pm} \ell^+ \ell^- \, \sla{E}_T,\;
    \ell^\pm jj \, \sla{E}_T \hbox{ and } \ell^+ \ell^- jj \;\;\;
\label{ppwz}
\end{eqnarray}
where $\ell$ and $\ell^\prime$ stand for electrons and muons and $j$ 
for jets. \medskip

In this work we present a {\sl model independent} analysis of the
observability of these signals. We express our results as function of
the relevant spin-1 boson effective couplings, mass and width.  For
$Z^\prime$ states these are the most sensitive channels which can give
information on its connection to EWSB. For $W^\prime$ they complement its
observation in WBF and provide further information on the $W^\prime$
properties.  The outline of this work is as follows: in
Sec.~\ref{sec:frame} we present the framework of our analysis and the
underlying assumptions in our model independent approach. In
Sec.~\ref{sec:signal} we describe the differentiating characteristics
between the signal and SM backgrounds contributing to the processes
(\ref{ppww}) and (\ref{ppwz}) and quantify those after imposing the
corresponding cuts to optimize the signal to background ratio. Finally,
in Sec.~\ref{sec:conclu} we present and discuss the obtainable
sensitivities in the different channels as a function of the relevant
parameters.

\section{Framework of our analyses}
\label{sec:frame}
From the phenomenological point of view, the study of processes
(\ref{ppww}) and (\ref{ppwz}) 
requires the knowledge of the couplings
of the new spin--1 states to the light quarks and to electroweak
vector boson pairs, in addition to their masses and widths.  \medskip

In this work, for the sake of concreteness, we assume that the couplings 
of the $Z^\prime$ and the $W^\prime$ to 
the light quarks and to gauge bosons have the same Lorentz structure as those
of the SM, as it is the case for Higgsless
models, but with rescaled strength. In Higgsless models, the
unitarity bound in $W^+W^- \rightarrow W^+W^-$ is saturated by the
exchange of the first Kaluza-Klein excitation (or $Z^\prime$) if its
couplings to electroweak gauge bosons is given
by~\cite{Birkedal:2004au}
\begin{equation}
{g_{Z^\prime WW}}_{max}=g_{ZWW}\, \frac{M_Z}{\sqrt{3}M_{Z^\prime}} 
\label{eq:gwwvmax}
\end{equation}
for a given $Z^\prime$ mass. Correspondingly, the saturation of unitarity in
the elastic scattering $W^\pm Z \to W^\pm Z$ leads to the
constraint~\cite{Birkedal:2004au}
\begin{equation}
{g_{W^\prime WZ}}_{max}
=g_{WWZ}\, \frac{M^2_Z}{\sqrt{3}
M_{W^\prime} M_W}  \; .
\label{eq:gwzvmax}
\end{equation}
These constraints imply an upper bound on the decay width of the
KK-bosons into SM gauge bosons.  In what follows we use 
${g_{W^\prime WZ}}_{max}$ and ${g_{Z^\prime WW}}_{max}$  simply as convenient
normalizations for the coupling of the spin-1 resonance to 
SM gauge bosons.
Moreover, in a generic model, the new
spin--1 states have further couplings to other particles, {\em
  e.g.}  $b$ or $t$ quarks, that contribute to their width. Therefore,
in this work we treat the $Z^\prime$ and $W^\prime$ widths as  free
parameters. With these assumptions it is possible to make a model
independent analysis of the observability of the relevant
processes. \medskip

Within this framework  the cross section for the processes (\ref{ppww})
and  (\ref{ppwz}) can be written in full generality as
\begin{eqnarray}
\sigma_{tot}= && \sigma_{SM}\; +\; 
\left(\frac{g_{V^\prime q\bar q}}{g_{Vq\bar q}} \frac{g_{V^\prime WV}}{
{g_{V^\prime WV}}_{max}}\right) 
\, \sigma_{int}(M_{V^\prime},\Gamma_{V^\prime})
\nonumber
\\
&&\;+\;
\left(\frac{g_{V^\prime q\bar q}}{g_{Vq\bar q}} \frac{g_{V^\prime WV}}{
{g_{V^\prime WV}}_{max}}\right)^2
\sigma_{V^\prime}(M_{V^\prime},\Gamma_{V^\prime}) 
\label{eq:sigmatot}
\end{eqnarray}
where for processes (\ref{ppww}) $V^\prime=Z^\prime$, 
$g_{V^\prime WV}\equiv g_{Z^\prime WW}$,
and $g_{Vq\bar q}\equiv g_{Zq\bar q}=g/c_W$.  For processes
(\ref{ppwz}) $V^\prime=W^\prime$, $g_{V^\prime WV}\equiv g_{W^\prime WZ}$, 
and $g_{Vq\bar  q}\equiv g_{Wq\bar q^\prime}=g/\sqrt{2}$.  
Here $g$ is the $SU(2)_L$
coupling constant and $c_W$ the cosine of the weak mixing angle.  We
notice that the final state $p p \to \ell^\pm jj \, \sla{E}_T$ can
receive contributions from both $Z^\prime$ and $W^\prime$ intermediate
states. When studying this channel we consider the sensitivity to
each of the contributions separately.  \medskip

In this approach, for each final state the analysis depends on three
parameters: the mass of the new spin-1 gauge boson, $M_{V^\prime}$, its
width, $\Gamma_{V^\prime}$, and the product of its couplings to light quarks
and to SM gauge bosons, $g_{V^\prime q\bar q}\; g_{V^\prime V V}$.  These
parameters are only subject to the constraint that for a given value
of product of the couplings of the new spin-1 boson and of its mass,
there is lower bound on its width, since
\begin{eqnarray}
&&\Gamma_{Z^\prime} \geq {\displaystyle \sum_{q=u,d}} \Gamma(Z^\prime \rightarrow q\bar q)
+\Gamma(Z^\prime \rightarrow W^+ W^-)  \label{eq:zwidmin}
\\
&&\Gamma_{W^{\prime +}} \geq \Gamma(W^{\prime +}\rightarrow u\bar d)
+\Gamma(W^{\prime +} \rightarrow W^+Z) \; .
\end{eqnarray}
Using the values  the partial widths (all in GeV):
\begin{eqnarray*}
&& \Gamma(Z^\prime \rightarrow u\bar u)=0.3 \, \left(\frac{M_{Z^\prime}}{M_Z}\right)\, 
\left(\frac{g_{Z^\prime q\bar q}}{g_{Zq\bar q}}\right)^2
\\
&& \Gamma(Z^\prime \rightarrow d\bar d)=0.38\, \left(\frac{M_{Z^\prime}}{M_Z}\right) \,
\left(\frac{g_{Z^\prime q\bar q}}{g_{Zq\bar q}}\right)^2
\\
&& \Gamma(Z^\prime \to W^+W^-)=0.028\, \left(\frac{M_{Z^\prime}}{M_Z}\right)^3 \,
\left(\frac{g_{Z^\prime WW}}{{g_{Z^\prime WW}}_{max}}
\right)^2 \\
&& \Gamma(W^{\prime +}\rightarrow q^\prime \bar q)=0.68 \, \left(\frac{M_{W^\prime}}{M_W}\right)\, 
\left(\frac{g_{W^\prime q\bar q}}{g_{Wq\bar q}}\right)^2
\\
&& \Gamma(W^{\prime +} \to W^+Z)=0.019\, \left(\frac{M_{W^\prime}}{M_W}\right)^3 \,
\left(\frac{g_{W^\prime WZ}}{{g_{W^\prime WZ}}_{max}}
\right)^2 \; , 
\label{eq:widths}
\end{eqnarray*}
it is possible to show that the minimum $V^\prime$ decay widths (in GeV)
are:
\begin{eqnarray}
&&\Gamma_{Z^\prime}\; >\;0.27 \, \left(\frac{g_{Z^\prime q\bar q}}{g_{Zq\bar q}}\right)\,
\left(\frac{g_{Z^\prime WW}}{{g_{Z^\prime WW}}_{max}}\right)\,
\,\left(\frac{M_{Z^\prime}}{M_Z}\right)^2 \; 
\; 
\label{eq:zcouplimit}
\\
&&\Gamma_{W^\prime}\; >\;0.40 \, \left(\frac{g_{W^\prime q\bar q}}{g_{Wq\bar q}}\right)\,
\left(\frac{g_{W^\prime WZ}}{{g_{W^\prime WZ}}_{max}}\right)\,
\,\left(\frac{M_{W^\prime}}{M_W}\right)^2 \;\; .
\label{eq:wcouplimit}
\end{eqnarray} 

We perform our parton--level analyses using the stand alone package
of MADGRAPH~\cite{madevent} supplemented by the new states and
interactions. We simulate experimental resolutions by smearing the
energies, but not directions, of all final state leptons with a
Gaussian error given by a resolution $\Delta E/E = 0.1/\sqrt{E} \oplus
0.01$ while for jets we assumed a resolution $\Delta E/E =
0.5/\sqrt{E} \oplus 0.03$, if $|\eta_j| \leq3$, and $\Delta E/E =
1/\sqrt{E} \oplus 0.07$, if $|\eta_j| >3$ ($E$ in GeV).  Furthermore
the lepton detection efficiency was taken to be $0.9$.  In our
calculations we use CTEQ6L parton distribution functions \cite{CTEQ6}
with renormalization and factorization scales $\mu_F^0 = \mu_R^0 =
\sqrt{({p^{\ell^+}_{T}}^2+{p^{\ell^-}_{T}}^2)/2}$ for pure leptonic final
states, while for the channels containing jets we employ $\mu_F^0 =
\mu_R^0 = \sqrt{({p^{j_2}_{T}}^2+{p^{j_1}_{T}}^2)/2}$. \medskip

\section{Signal and backgrounds}
\label{sec:signal}

In this section we describe the main features of the $Z^\prime$ and $W^\prime$
signals for the different final states, presenting the main SM
backgrounds and discuss possible cuts to enhance the signal to 
backgrounds ratios.

\subsection{$\mathbf{p p \to Z^\prime \to W^+ W^- \to \ell^+ \ell^{\prime -} 
\, \sla{E}_T}$}
\label{subsec:ZPl}
We start discussing the observability of the $Z^\prime$ signal in the leptonic
$W^\pm$ decay channels with both equal and different flavor leptons
in the final state. In the case of different flavor leptons, the $Z^\prime$
signal in process (\ref{ppww}) possesses SM backgrounds coming from
the production of $W^+ W^-$ pairs, as well as, from $t \bar{t}$ pairs
where the top quarks decay semi-leptonically. \medskip

The starting cuts are meant to ensure the detection and isolation
of the final partons plus a minimum transverse momentum 
since $Z^\prime$ has a mass in excess of a few hundred GeV~\cite{teva} 
and it is expected to lead to hard leptons  
\begin{equation}
| \eta_\ell| < 2.5 \hbox{ , }\;\;
\Delta R_{\ell\ell} > 0.4 
\;\;
\hbox{ and } 
 \;\;
p_T^\ell > 50 \; \hbox{ GeV} \;.
\label{cutsww1}
\end{equation}

The presence of two neutrinos in the final state renders impossible
the complete reconstruction of the event. Nevertheless
the signal can still be partially characterized with the use of the 
transverse invariant mass variable 
\begin{eqnarray}
  M_T^{WW} 
=&& \biggl[ \left( \sqrt{(p_T^{\ell^+\ell^{\prime -}})^2 + m^2_{\ell^+ \ell^{\prime -}}}   
              + \sqrt{\sla{p_T}^2 + m^2_{\ell^+\ell^{\prime -} }} \right)^2 \biggr .
\nonumber \\ 
&&\biggl . - (\vec{p}_T^{~\ell^+\ell^{\prime -}} + \vec{\sla{p_T}}  )^2 \biggr]^{1/2} 
\label{eq:mtww}
\end{eqnarray}
where $\vec{\sla{p_T}}$ is the missing transverse momentum vector,
$\vec{p}_T^{~\ell^+\ell^{\prime -}}$ is the transverse momentum of the pair
$\ell^+ \ell^{\prime -}$ and $m_{\ell^+\ell^{\prime -}}$ is the $\ell^+ 
\ell^{\prime -}$ invariant mass. \medskip

\begin{figure}[t]
\includegraphics[width=3.5in]{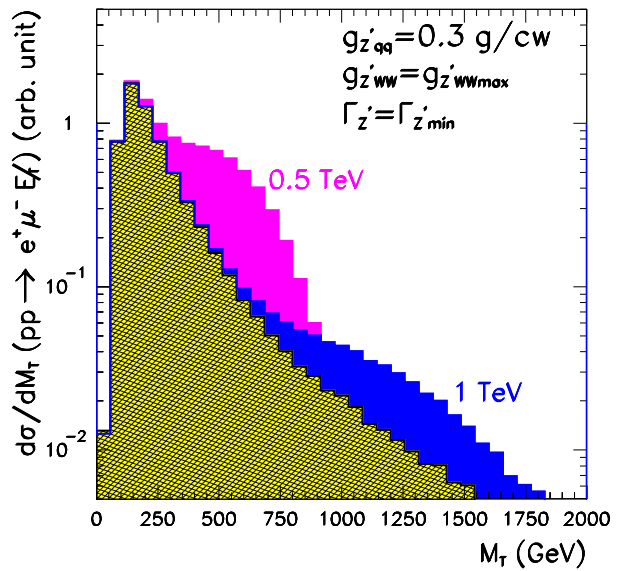}
\caption{Transverse mass distribution for the $pp\rightarrow e^+ \mu^-
  \, \sla{E}_T$ final state for different values of $M_{Z^\prime}$. The widths
are chosen to be the minimum values compatible with Eq.~(\ref{eq:zwidmin}),
$\Gamma_{min}=4.7,\; 37$ GeV for 0.5 and 1 TeV, respectively.
The hatched
  area stands for the SM background contribution.}
\label{fig:mtdist}
\end{figure}

We show in Fig.~\ref{fig:mtdist} the characteristic invariant mass
distribution expected for several values of $M_{Z^\prime}$.  The SM
prediction is represented by the solid light grey (yellow) histogram
while the darker areas represent the $Z^\prime$ expected signal for $M_{Z^\prime}
= 0.5$ and 1 TeV as indicated. As seen in this figure, the
characteristic peak associated with the production of a resonance is
substantially broadened due to the incomplete event reconstruction,
however, the $Z^\prime$ production is still signaled by an excess of events
in the region
\begin{equation}
  M_T^{WW} 
>\frac{M_{Z^\prime}}{2}  \;.
\label{eq:mtcut}
\end{equation}
 
A sizable contribution contribution to the background arises from the 
SM $t \bar{t}$ production which leads to a $W^+ W^-$ pair on the final
state accompanied by two b--jets. This background can be efficiently
reduced by vetoing the presence of additional jets with
\begin{equation}
    | \eta_j | < 3   \;\;\;\; \hbox{ and } \;\;\; p_T^j > 20 \;\hbox{ GeV.}
\label{eq:veto}
\end{equation}
The probability of a QCD (electroweak) event to survive such a
central jet veto has been analyzed for various processes in
Ref.~\cite{rainth}. Moreover, at the high--luminosity run of the LHC
there will be more than one interaction per bunch crossing,
consequently there is a probability of detecting an extra jet in the
gap region due to pile--up.  In Ref.~\cite{atlas} it was estimated
that survival probability due to pile--up is 0.75 for a threshold cut
of $p_T=20$ GeV.  Taking into account these two effects we included in
our analysis veto survival probabilities
\begin{equation}
P_{\rm surv}^{\rm EW}= 0.56 \;\;\;\;\;\; , \;\;\;\;\;\;
P_{\rm surv}^{\rm QCD}=0.23 \;\; .
\label{eq:psurv}
\end{equation}

\begin{figure}[t]
\includegraphics[width=3.5in]{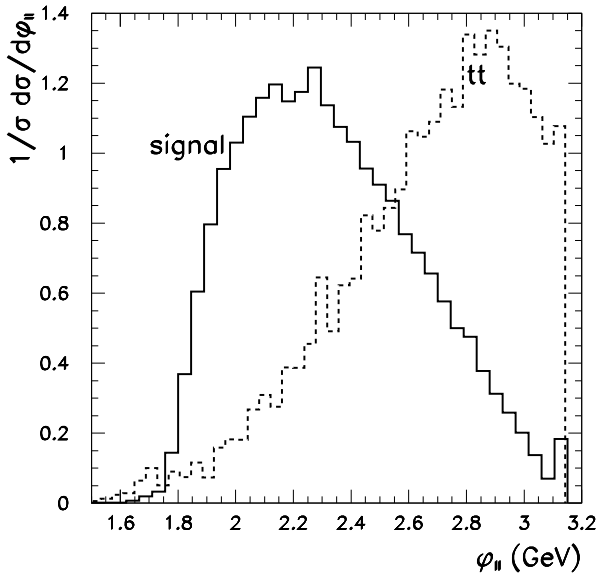}
\caption{Distribution of the azimuthal angular separation of the charged 
leptons  for the  $\ell^+ \ell^{\prime -} \,\sla{E}_T$ final state.
The solid histogram stands for the signal distribution for $M_{Z^\prime}=0.5$ TeV
with $\Gamma_{Z^\prime}/M_{Z^\prime}$=0.1,  while the dashed one
  represents the SM $t \bar{t}$ background.}
\label{fig:delphill}
\end{figure}

Further reduction of $t \bar{t}$ background can be achieved by use of
the difference of the azimuthal angular distributions of the final
leptons.  This behavior is illustrated in Fig.~\ref{fig:delphill}
where we show the azimuthal angular separation of the charged leptons
for both signal and $t \bar{t}$ background.  Consequently, one can 
also  require
\begin{equation}
\Delta\phi_{\ell \ell^\prime} < 2.5  \;\; .
\label{eq:phillc}
\end{equation}

In the case that the final state leptons in reaction (\ref{ppww}) have
the same flavor ($ee$ or $\mu\mu$), there are additional backgrounds
originating from Drell-Yan lepton pair production, as well as $ZZ$
pair production. In this case, the cuts (\ref{cutsww1}), (\ref{eq:mtcut}), 
(\ref{eq:veto}) and (\ref{eq:phillc}) are supplemented by
\begin{equation}
     \sla{E}_T > 50 \; \hbox{ GeV}
\label{etcut}
\end{equation}
in order to suppress the Drell-Yan pair production. Furthermore 
\begin{equation}
      m_{\ell^+ \ell^-} > 100 \; \hbox{ GeV.}
\label{mllcut}
\end{equation}
is required to reduce the $ZZ$ background.

Table \ref{tab:ww} contains the cross sections for the different contributions
to the SM backgrounds after all the cuts described above, as well as
for the signal, normalized as in Eq.~(\ref{eq:sigmatot}), for
$M_{Z^\prime}=0.5,~ 1,$ and $ 1.5$ TeV and for an illustrative value of the
width $\Gamma_{Z^\prime}=0.05 M_{Z^\prime}$.  Once the cuts are imposed, the
interference term $\sigma_{int}$ is negligibly small for all values of
$Z^\prime$ masses and widths considered.

\begin{table}[htb]
 \begin{tabular}{||c|c|c|c||}
\hline
\hline
$M_{Z^\prime}$ (TeV)   &  $\sigma_{SM}^{EW}$ (fb)  & 
 $\sigma_{SM}^{t\bar t}$ (fb) & $\sigma_{Z^\prime}$ (fb) \\ 
\hline 
0.5  &  184 &  10.8 & 793 \\  
1.0 & 92.5 &   1.88 & 361 \\
1.5  &  52 & 0 & 158 \\
\hline
\hline
 \end{tabular}
 \caption{Signal and background cross sections in fb for 
   the  $\ell^+ \ell^{\prime -} \,\sla{E}_T$ final state
   for all possible lepton combinations with either electrons
   or muons. For the final state with 
   different flavor leptons the cuts (\ref{cutsww1}) 
   to (\ref{eq:phillc}) are applied while for same 
   flavor leptons the cuts (\ref{cutsww1}) to (\ref{mllcut}) are imposed .
   These results do not include the lepton detection efficiencies
   nor the gap survival probabilities.
   For the signal the results are show for an illustrative value of the width 
   $\Gamma_{Z^\prime}=0.05 M_{Z^\prime}$.}
 \label{tab:ww}
\end{table}

\subsection{$\mathbf{p p \to Z^\prime \to W^+ W^- \to \ell^\pm jj \,\sla{E}_T}$}
\label{subsec:zpjjlnu}

A sharper $Z^\prime$ signal can be obtained by the study of the channel where one
$W$ decays leptonically while the other decays hadronically. This
final state can be reconstructed, up to a twofold ambiguity on the
neutrino longitudinal momentum, exhibiting the characteristic peak of
a resonance production.  On the other hand this channel possesses
large QCD backgrounds due to the $Wjj$ production, as well as, $Zjj$
with the $Z$ decaying leptonically and one of the leptons being
missed.  $t \bar{t}$ production also contribute to the background,
however, presenting additional $b$ jets. \medskip

Again the starting cuts are meant to ensure the detection and isolation
of the final partons: 
\begin{equation}
  | \eta_\ell | < 2.5  \;\;\;\;,\;\;\;\; 
| \eta_j| < 3 \;\;\;\;\hbox{and}\;\;\;\;
\Delta R_{jj(j\ell)} > 0.4 \; .
\label{jjlnu:c1}
\end{equation}

\begin{figure}[t]
\includegraphics[width=3.5in]{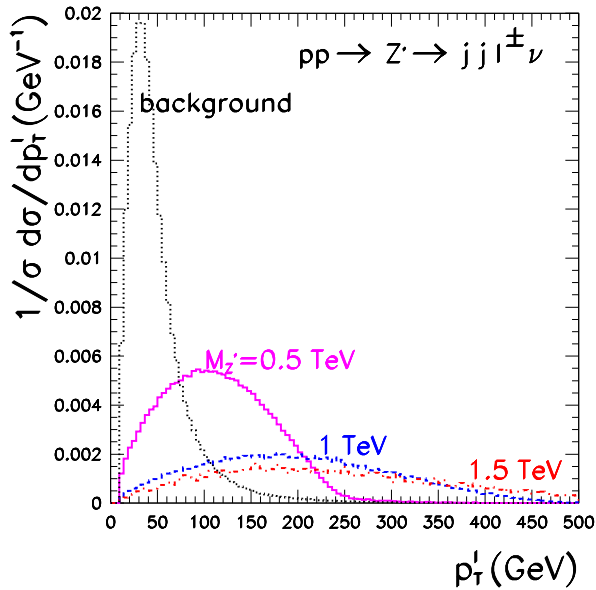}
\caption{
Transverse momentum distribution for the charge lepton in 
the $\ell^\pm jj \,\sla{E}_T$ final state. The dotted histogram 
corresponds to the expected distribution for the background 
after the cuts in Eq.(\ref{jjlnu:c1}) while the solid, dashed, and
dot-dashed line correspond to the expected distribution for the 
signal with $M_{Z^\prime}=0.5$, 1 and 1.5 TeV, respectively and for  
$\Gamma_{Z^\prime}=0.05 M_{Z^\prime}$.}
\label{fig:ptl}
\end{figure}

The large QCD background can be reduced by making use of the
characteristic harder transverse momenta of charged lepton and the
jets in the $Z^\prime$ signal.  As illustration we show in
Fig.~\ref{fig:ptl} the $p_T^\ell$ distribution for both the sum of the
backgrounds as well as for the signal for several values of $M_{Z^\prime}$.
Thus, the signal to background ratio can be improved by requiring the
lepton and jets to be energetic. We find the following suitable
set of variables:
\begin{equation}
\begin{array}{ll}
p_T^\ell > p_T^{\ell\, {\rm min}}  \;\;\;\;\;\; &\hbox{ , } \;\;\;\;\;
p_T^j > p_T^{j {\rm min}} \; 
\\
&
\\
p_T^j|_{\rm max} > p_T^{\rm max}
\;\;\;\;\;\; &\hbox{ , } \;\;\;\;\;
\sla{E}_T > \sla{E}_T^{\rm min}  \; ,
\end{array}
\label{wzjjl:c2}
\end{equation}
where we denoted the jet largest transverse
momentum as $p_T^j|_{\rm max}$.   The optimum cuts depend on the $Z^\prime$ 
mass as given in Table~\ref{tab:cutsjjlnu}.

The identification of the
hadronically decaying $W$ is accomplished requiring that the dijet
invariant mass is compatible with the $W$ mass, {\em i.e.}
\begin{equation}
| M_{jj} - M_W | < 10 \; \hbox{ GeV.}
\label{wzjjl:c3}
\end{equation}

As discussed in Sec.~\ref{subsec:ZPl}, the $t \bar{t}$ background can be 
further reduced by vetoing
additional jets in the central region through the requirement
(\ref{eq:veto}) which we assumed to lead to the corresponding veto
survival probabilities (\ref{eq:psurv}). This veto also suppresses the
QCD processes since QCD events presents a higher probability of
emitting additional jets~\cite{rainth}.

\medskip

\begin{table}[htb]
\begin{tabular}{||c|c|c|c||}
\hline
\hline
$M_{Z^\prime}$ (TeV) & 0.5 & 1.0 & 1.5\\
\hline 
$p_T^{\ell\, {\rm min}}$ (GeV) & 75 & 100 & 100 \\
$p_T^{j\, {\rm min}}$ (GeV) & 60 & 75 & 100 \\
$p_T^{\rm max}$ (GeV) & 110 & 75 & 100 \\
$ \sla{E}_T^{\rm min}$ (GeV) & 50 & 75 & 75 \\
$\delta$ (GeV) & 50 & 100 & 200 \\
\hline \hline 
\end{tabular}
\caption{Cuts for 
$p p \to Z^\prime \to W^+ W^- \to \ell^\pm jj \,\sla{E}_T$ as a function of the new 
resonance mass.}
\label{tab:cutsjjlnu}
\end{table}

\begin{figure}[t]
\includegraphics[width=3.5in]{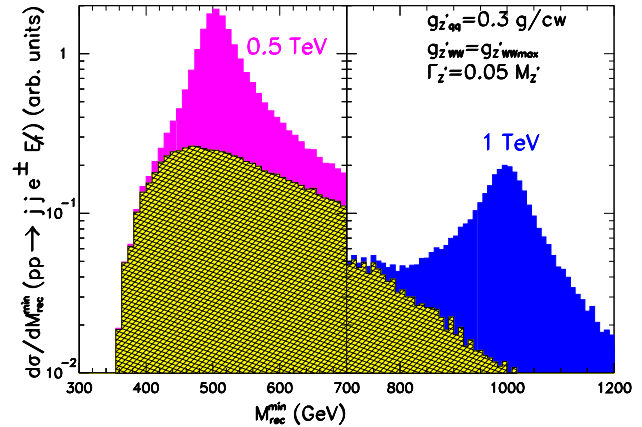}
\caption{In the left (right) panel we present the reconstructed
  minimum invariant mass distribution for the $pp\rightarrow j j
  \ell^\pm \,\sla{E}_T$ final state for $M_{Z^\prime} = 0.5$ TeV (1 TeV)
  over the corresponding SM backgrounds (hatched area) after the cuts
  in Eqs.~(\ref{jjlnu:c1})--(\ref{wzjjl:c3}). Despite the arbitrary
  overall normalization the figure reflects the relative size of the two 
resonance signals (and of their corresponding backgrounds).}
\label{fig:mtdist_jjln}
\end{figure}

As mentioned above, the presence of just one neutrino in the final
state permits the reconstruction of its momentum by imposing the
transverse momentum conservation and requiring that the invariant mass
of the neutrino--$\ell^\pm$ pair is $M_W$; this procedure leads a
twofold ambiguity on neutrino longitudinal momentum. This opens up the
possibility of reconstructing the $Z^\prime$ Breit--Wigner profile to better
pinpoint this state.  Using the two possibilities for the neutrino
momentum we evaluated the invariant mass of the $\ell^\pm jj
\,\sla{E}_T$ final state.  In Fig.~\ref{fig:mtdist_jjln} we show the
reconstructed invariant mass distribution using the minimum of the
two  $WW$ reconstructed masses, $M^{\rm min}_{\rm rec}$, after the cuts in
Eqs.~(\ref{jjlnu:c1})--(\ref{wzjjl:c3}). \medskip

For any of the widths considered in the analysis, the $Z^\prime$ peak can 
be efficiently observed by requiring:
\begin{equation}
| M^{\rm min}_{\rm rec} - M_{Z^\prime} | < \delta \; ,
\label{wzjjl:c4}
\end{equation}
where $\delta$ is chosen to guarantee that for a given mass and for
the range of widths considered most of the signal is within this
invariant mass window. The chosen values of $\delta$ are given in
Table \ref{tab:cutsjjlnu}. \medskip

\begin{figure}[t]
\includegraphics[width=3.5in]{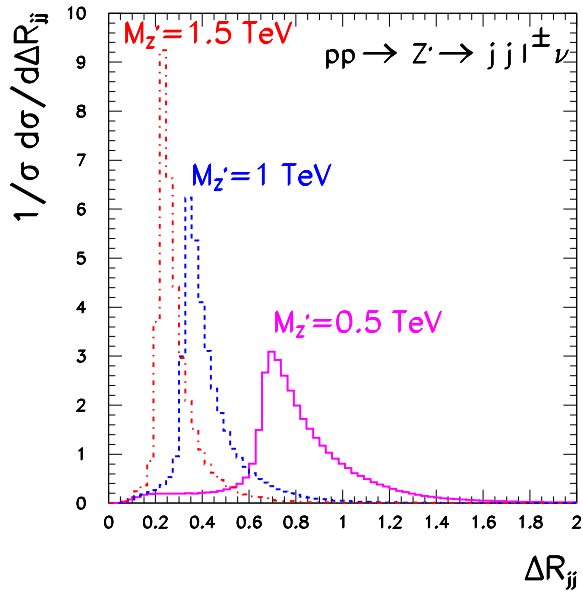}
\caption{$\Delta R_{jj}$ distribution in 
the $\ell^\pm jj \,\sla{E}_T$ final state for different 
$M_{Z^\prime}$ signals as labeled in the figures. In all cases 
$\Gamma_{Z^\prime}=0.05 M_{Z^\prime}$.}
\label{fig:dr}
\end{figure}

We present in Table \ref{tab:zprime3} the signal and background cross
sections for the $\ell^\pm jj \,\sla{E}_T$ channel after applying the
cuts (\ref{jjlnu:c1}) to (\ref{wzjjl:c4}) \footnote{Again, once the
  cuts are imposed, the interference term $\sigma_{int}$ is negligibly
  small for all values of $Z^\prime$ masses and widths considered. }.  From
this table, we can see that the cross sections are drastically reduced
for $Z^\prime$ masses in excess of $\simeq 1$ TeV.  In this region of the
parameter space the $W$'s are very energetic, thus, their decay
products are highly collimated not passing the isolation cuts in
(\ref{jjlnu:c1}).  This is illustrated in Fig.~\ref{fig:dr} where we
show the $\Delta R_{jj}$ distribution expected from the $Z^\prime$ signal
for different values of $M_{Z^\prime}$ when only the rapidity cuts in
Eq.\ (\ref{jjlnu:c1}) are imposed. As seen in the figure, already for
$M_{Z^\prime}=1.5$ TeV a sizeable fraction of the signal leads to very
collimated jets, which results into the signal reduction once the
isolation cuts are imposed. \medskip

In order to overcome this suppression one can consider the possibility
of the final state $\ell^\pm j \,\sla{E}_T$, as discussed in
Ref.~\cite{han}.  The realistic evaluation of the attainable level of
rejection of the QCD background for this case requires a devoted study
including details of the detector simulation, which is beyond the
scope of this work.  We therefore keep the isolation cut which may
render our conclusions of observability of the heavy resonances
conservative.
 
\begin{table}[htb]
\begin{center}
\begin{tabular}{||c|c|c|c|c||} 
\hline
\hline
$M_{Z^\prime}$ (TeV) &
  $\sigma_{SM}^{EW}$ (fb)  & $\sigma_{SM}^{QCD}$ (fb) & 
$\sigma_{SM}^{t\bar{t}}$ (fb) & $\sigma_{Z^\prime}$ (fb)\\
\hline
0.5  & 17.6  & 208 & 19.2& 1232\\
1.0  & 5.6  & 46 & 5.6 & 270\\
1.5  & 1.5 & 12 & 1.8& 9.5  \\
\hline
\hline
\end{tabular}
\end{center}
\caption{Signal and background cross sections in fb 
  for  the final state   $\ell^\pm jj \,\sla{E}_T$ 
  after imposing the cuts
  (\ref{jjlnu:c1}) to (\ref{wzjjl:c4}) 
  summed over $\ell^\pm$ being an electron 
  or a muon. 
  These results do not include the jet and lepton reconstruction efficiencies.
  For the signal the results are show for an illustrative value of the width 
  $\Gamma_{Z^\prime}=0.05 M_{Z^\prime}$.}
\label{tab:zprime3}
\end{table}

\subsection{$\mathbf{p p \to W^\prime  \to W^\pm Z \to \ell^{\prime \pm}
\ell^+ \ell^- \,  \sla{E}_T}$}

In this channel, the $W^\prime$ production is characterized by the presence
of three charged leptons ($e^\pm$ or $\mu^\pm$) and missing transverse
energy. The primary SM background to this signal is the $W^\pm Z$ and
$ZZ$ productions, where one of the final state leptons evades
detection in the $ZZ$ case. The $t \bar{t}$ pair production also
contributes to the SM background, however, since it requires that one
of the isolated leptons originates from a $b$ quark semi-leptonic
decay it is very suppressed. \medskip

We start with the minimum cuts to ensure the detection and isolation
of the final leptons: 
\begin{equation}
|\eta_\ell| < 2.5 \;\;\;,\;\;\;
\Delta R_{\ell \ell}> 0.4 \;\;\;\hbox{and}\;\;\; 
p_{T}^\ell > 10 \hbox{ GeV.}
\label{basiccuts}
\end{equation}
In the search for a lighter resonance, {\em i.e.} $M_{W^\prime} = 0.5$
TeV, a further cut is needed to tame the SM backgrounds, demanding
that the hardest observed lepton has transverse momentum in excess of
120 GeV. \medskip

In the signal the final state contains a pair of
same flavor opposite charge (SFOC) leptons compatible with the production of
a $Z$. Thus the signal gets enhanced by requiring 
\begin{equation}
\left | M_{\ell\ell}^{\rm SFOC}   - M_Z \right | < 20 \hbox{ GeV.}
\label{mllmz}
\end{equation}

As in Sec.~\ref{subsec:zpjjlnu},  
the presence of just one final state neutrino in process
(\ref{ppwz}) allows for the reconstruction of the neutrino momentum.
Consequently it is possible to evaluate the total $\ell\ell\ell^\prime \nu$
invariant mass up to a twofold ambiguity. Thus one can enhance the signal
over SM backgrounds by requiring that the total reconstructed
invariant mass satisfies
\begin{eqnarray}
\left | M_{\rm rec}^{\rm min}- M_{W^\prime} \right | < \delta \; ,
\label{mwzcut}
\end{eqnarray}
where $M_{\rm rec}^{\rm min}$ is the smaller of the two solutions for
the reconstructed invariant mass. We chose the size of the window
around $M_{W^\prime}$ as given in Table~\ref{tab:cutsjjlnu} with the
implicit understanding that $M_{Z^\prime}$ should be taken as the
$W^\prime$ mass.  \medskip

We present in Table~\ref{tab:wprimelep} the signal and SM cross
sections after cuts (\ref{basiccuts})--(\ref{mwzcut}).  The
interference term $\sigma_{int}$ is negligibly small for all values of
$W^\prime$ masses and widths considered.  As we can see, this channel
presents a small SM background due to the reduced leptonic branching
ratio of the $WZ$ final state. However, the signal cross section is
depleted as well.  As discussed in the previous subsection, the
collimation of the $Z$ decay products reduces the signal for large
$M_{W^\prime}$ due to the requirement of the lepton isolation cuts.
\medskip

\begin{table}[htb]
\begin{center}
\begin{tabular}{||c|c|c||} 
\hline
\hline
$M_{W^\prime}$ (TeV) 
 & $\sigma_{SM}$ (fb)  
&  $\sigma_{W^\prime}$ (fb) \\
\hline
0.5 & 0.87 & 297 \\
1.0 & 0.14   & 85  \\
1.5 &  0.01 & 6.4 \\
\hline
\hline
\end{tabular}
\end{center}
\caption{Signal and background cross sections for the 
  $\ell^{\prime \pm} \ell^+ \ell^- \,  \sla{E}_T$ final state
  after imposing cuts (\ref{basiccuts})--(\ref{mwzcut}). 
  Contributions from all trilepton combinations with electrons and muons 
  are included.    Backgrounds include both irreducible and reducible 
  contributions plus  the top backgrounds. The lepton detection efficiencies  
  are not included. The signal cross section is given for 
  $\Gamma_{W^\prime}=0.05 M_{W^\prime}$}. 
\label{tab:wprimelep}
\end{table}

\subsection{$\mathbf{p p \to W^\prime \to W^\pm Z \to \ell^\pm jj  \,  \sla{E}_T}$}

The final state $\ell^\pm jj \,\sla{E}_T$ can also be used to study
$W^\prime$ production, in addition to $Z^\prime$ searches described in
Sec.~\ref{subsec:zpjjlnu}. Since the $W$ and $Z$ masses are relatively
close, many events might be classified as $Z^\prime$ and $W^\prime$
productions if the new vector resonances possess similar masses. In
our analyses we assumed that these new states are not degenerated so
we did not add their contributions to this final state. \medskip

The $W^\prime$ and $Z^\prime$ signals are similar for this final state, therefore,
the same cuts (\ref{jjlnu:c1}) to (\ref{wzjjl:c4}) can be applied with the
obvious change of $M_W$ to $M_Z$ in (\ref{wzjjl:c3}) and of $M_{Z^\prime}$
by $M_{W^\prime}$ in (\ref{wzjjl:c4}).  Also to efficiently suppress the $t
\bar{t}$ and QCD backgrounds the vetoing of additional jets
in the central region (\ref{eq:veto}) is imposed with 
the corresponding veto survival probabilities
(\ref{eq:psurv}).  \medskip

Table~\ref{tab:wprime3} shows the signal and background cross sections
for the $W^\prime$ search in the $\ell^\pm jj \,\sla{E}_T$ channel.
Comparison with Table \ref{tab:wprimelep} shows that in this channel 
the signal has a higher statistics than the purely leptonic
mode due to the larger $Z$ hadronic branching ratio. However, as
seen in the table, even after all the cuts imposed,  
this channel still suffers a large QCD background.

\begin{table}[htb]
\begin{center}
\begin{tabular}{||c|c|c|c|c||} 
\hline
\hline
$M_{W^\prime}$ (TeV) & $\sigma_{SM}^{EW}$ (fb)  
&  $\sigma_{SM}^{QCD}$ (fb) & $\sigma_{SM}^{t\bar{t}}$ (fb) 
& $\sigma_{W^\prime}$ (fb)\\
\hline
0.5  & 13.2  & 210. & 11.4 & 1354\\
1.0  & 4.0  & 50. & 4.4 & 673\\
1.5  & 1.5 & 14. & 0.88 & 32\\
\hline
\hline
\end{tabular}
\end{center}
\caption{ $W^\prime$ signal and SM background cross sections for  the final state 
  $\ell^\pm jj \, \sla{E}_T$ as a function of the $W^\prime$ mass.
  Results are summed over $\ell^\pm$ being an electron 
  or a muon and  do not include the jet and lepton reconstruction efficiencies.
  For the signal the results are show for an illustrative value of the width 
  $\Gamma_{W^\prime}=0.05 M_{W^\prime}$.}
\label{tab:wprime3}
\end{table}

\subsection{$\mathbf{p p \to W^\prime \to W^\pm Z \to \ell^+ \ell^- jj} $}

Finally we study the final state $\ell^+ \ell^- jj$ for which the 
full reconstruction
of the $W^\prime$ signal as well as the $Z$ and $W$ intermediate states is possible.
This, as shown below, allows for a very efficient reduction of the SM
backgrounds. \medskip

The centrality, isolation and momentum cuts for the final partons 
for this channel are:
\begin{equation}
  | \eta_\ell | < 2.5 \;\;\;\;,\;\;\;\;
 | \eta_j| < 3 \;\;\;\;\hbox{and}\;\;\;\;
\Delta R_{jj(j\ell)(\ell\ell)} > 0.4 \; .
\label{jjll:c1}
\end{equation}
\begin{eqnarray}
&&p_T^\ell > 50,~(100),~[100]\; {\rm GeV}  \;\;\hbox{ , } \nonumber \\
&& \label{jjll:c2}
\\
&&p_T^j > 70,~(100),~[100] \;    {\rm GeV} \,  
\nonumber\; 
\end {eqnarray}
for $M_{W^\prime}=0.5,~(1.0),~[1.5]$ TeV. \medskip

The identification of the 
hadronically decaying $W$ and leptonically decaying $Z$ is
obtained by requiring:
\begin{eqnarray}
&&| M_{jj} - M_W | < 10 \; \hbox{ GeV,} \nonumber \\ 
&& \label{jjll:c3}
\\
&&| M_{\ell^+\ell^-} - M_Z | < 10 \; \hbox{ GeV} . 
\nonumber
\end{eqnarray}

\begin{figure}[t]
\includegraphics[width=3.5in]{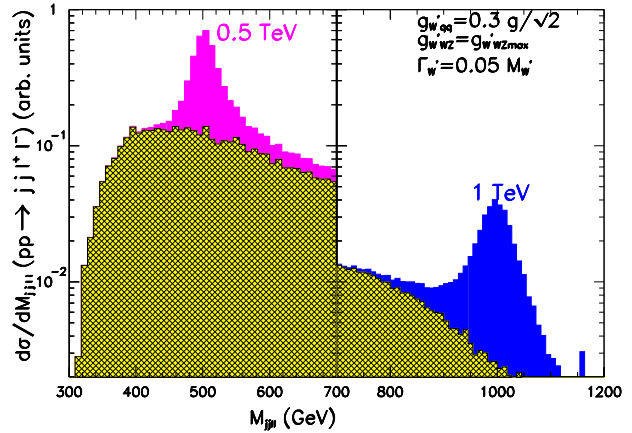}
\caption{The left (right) panel contains the reconstructed invariant
  mass distribution for the $pp\rightarrow j j \ell^+ \ell^-$ final
  state and for $M_{Z^\prime}= 0.5$ (1) TeV over their corresponding
  backgrounds, depicted as a hatched area, after the cuts given by
  Eqs.~(\ref{jjll:c1})--(\ref{jjll:c3}). The relative size of
  the signals and backgrounds between the two resonances is maintained 
  despite the arbitrary  overall normalization.}
\label{fig:mtdist_jjll}
\end{figure}

In Fig.~\ref{fig:mtdist_jjll} we show the
reconstructed invariant mass distribution $M_{jj\ell^+\ell^-}$ after 
the  cuts in Eqs.~(\ref{jjll:c1})--(\ref{jjll:c3}). 
As seen in this figure, in this channel the $W^\prime$ invariant mass can 
be well reconstructed. We consequently compute the signal
cross section for $M_{W^\prime}=0.5 \;(1)\; [1.5]$ TeV by requiring 
\begin{eqnarray}
\left | M_{jj\ell^+ \ell^-}- M_{W^\prime} \right | < 
50,~ (100),~  [200] \, {\rm GeV.}  \; 
\label{mjjllcut}
\end{eqnarray}

We give in Table \ref{tab:wprime4} the signal and SM cross sections after 
cuts (\ref{jjll:c1})--(\ref{mjjllcut}).  The $t\bar{t}$ cross section is 
reduced to a negligible size after these cuts are imposed.
The interference term $\sigma_{int}$ is negligibly small for all values
of $W^\prime$ masses and widths considered. 

\begin{table}[htb]
\begin{center}
\begin{tabular}{||c|c|c|c||} 
\hline
\hline
$M_{W^\prime}$ (TeV) & $\sigma_{SM}^{EW}$ (fb)  
&  $\sigma_{SM}^{QCD}$ (fb) & $\sigma_{W^\prime}$ (fb)\\
\hline
0.5  &  12.2  & 17.2 &  380\\
1.0  &  0.14 &  0.76 &  51 \\
1.5  &  0.03 &  0.12  &  3.3 \\
\hline
\hline
\end{tabular}
\end{center}
\caption{
$W^\prime$ signal and SM background cross sections for  the final state 
  $\ell^+ \ell^- jj $ as a function of the $W^\prime$ mass.
Results are summed over $\ell^\pm$ being an electron 
or a muon and  do not include the jet and lepton reconstruction efficiencies.
For the signal the results are show for an illustrative value of the width 
$\Gamma_{W^\prime}=0.05 M_{W^\prime}$.}
\label{tab:wprime4}
\end{table}

As in the previous section,  the collimation of the $Z$  and $W$ decay
products reduces the signal for large $M_{W^\prime}$ due to the requirement 
of the lepton and jet isolation cuts imposed  to ensure efficient 
reconstruction of the $Z$  and $W$ invariant masses. 

\begin{figure}[t]
\includegraphics[width=3.5in]{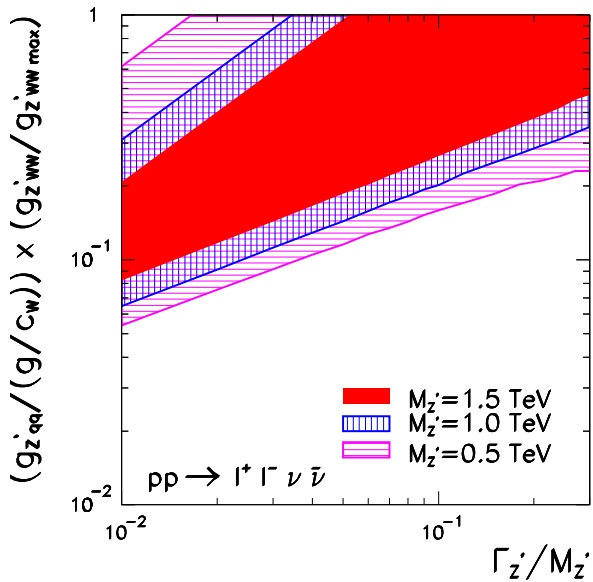}
\caption{The filled regions are the ranges of the parameters for
  observation of a $Z'$ with mass $M_{Z^\prime}=0.5$, 1, and 1.5 TeV with at
  least 5$\sigma$ significance in the reaction $pp \to Z^\prime \to
  W^+ W^- \to \ell^+ \ell^- \,\sla{E}_T$.}
\label{fig:boundsv0lep}
\end{figure}
\begin{figure}[t]
\includegraphics[width=3.5in]{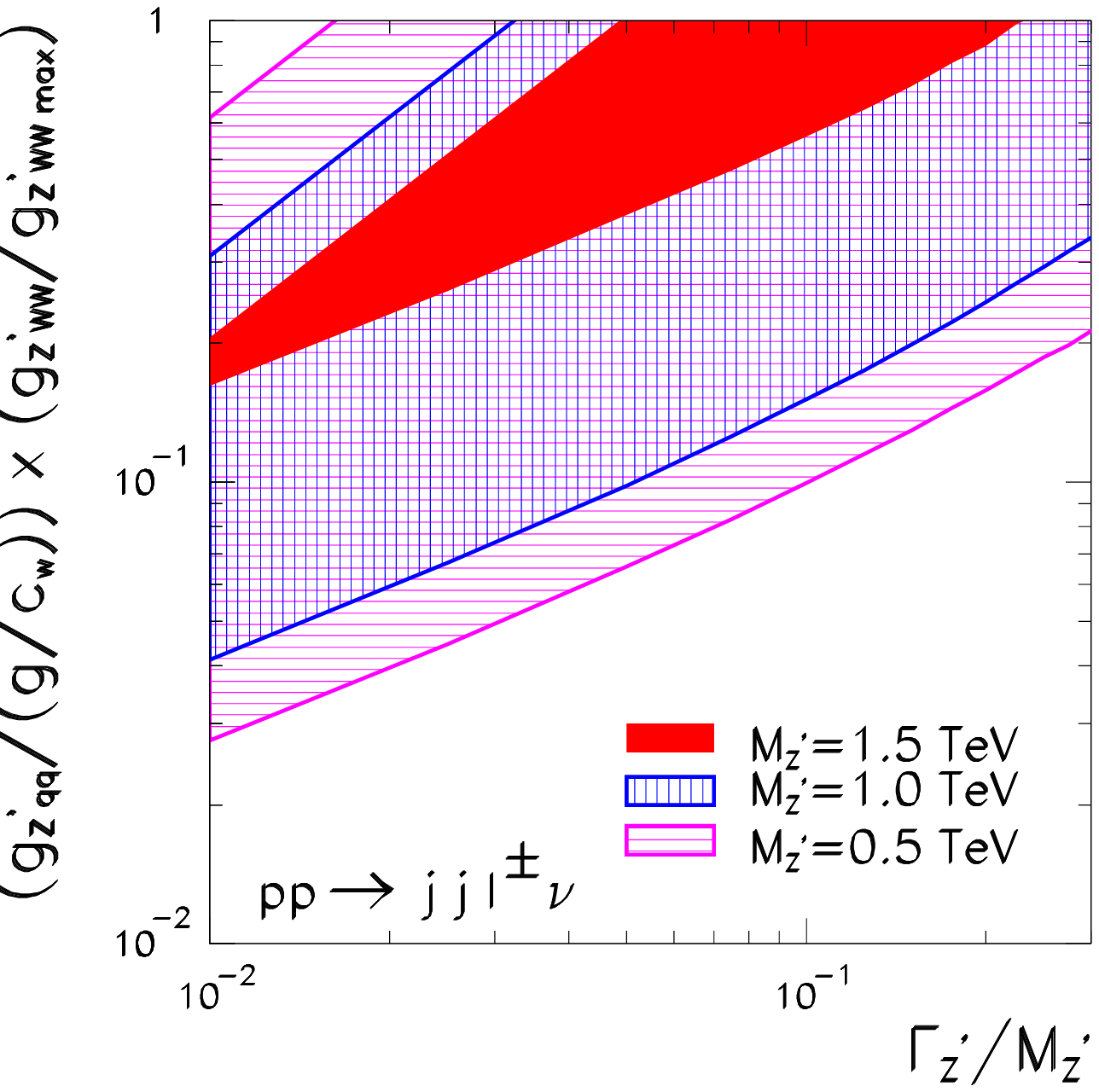}
\caption{Same as Fig.~\ref{fig:boundsv0lep} for the process $p p \to
  Z^\prime \to W^+ W^- \to \ell^\pm jj \,\sla{E}_T$.}
\label{fig:boundsv0nljj}
\end{figure}
\begin{figure}[t]
\includegraphics[width=3.5in]{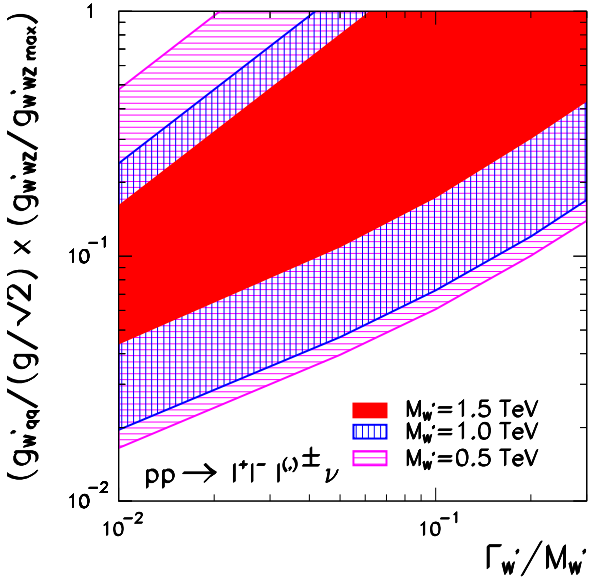}
\caption{The filled regions are the ranges of the parameters for
  observation of a $W^\prime$ with mass $M_{W^\prime}=0.5$, 1, and 1.5 TeV with at
  least 5$\sigma$ significance in the channel $pp \to W^\prime \to
  W^\pm Z \to \ell^+ \ell^- \ell^\pm \,\sla{E}_T$.}
\label{fig:boundswplep}
\end{figure}

\begin{figure}[t]
\includegraphics[width=3.5in]{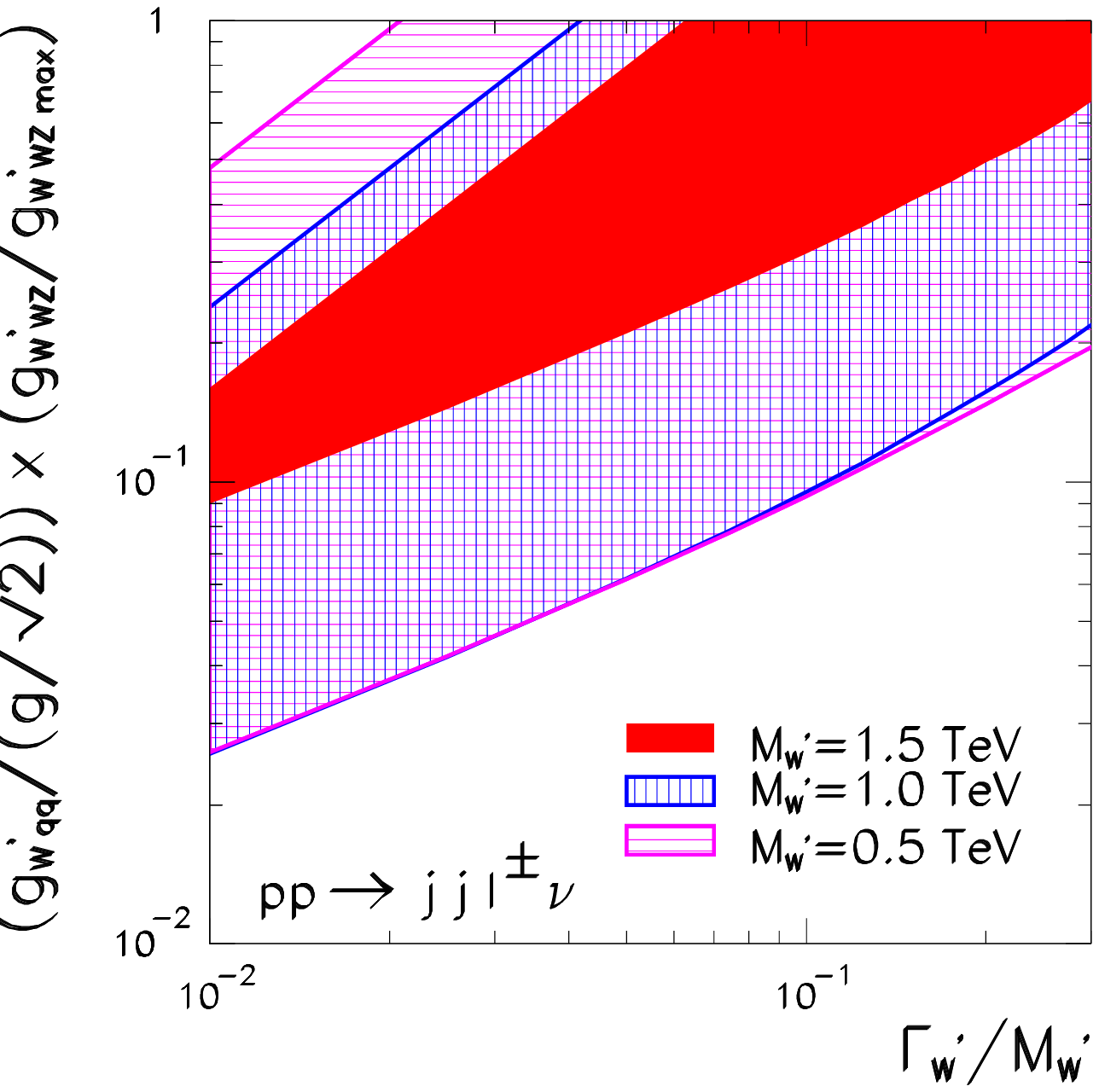}
\caption{Same as Fig.~\ref{fig:boundswplep} for the reaction $p p \to
  W^\prime \to W^\pm Z \to \ell^\pm jj \,\sla{E}_T$.}
\label{fig:boundswpnljj}
\end{figure}

\begin{figure}[t]
\includegraphics[width=3.5in]{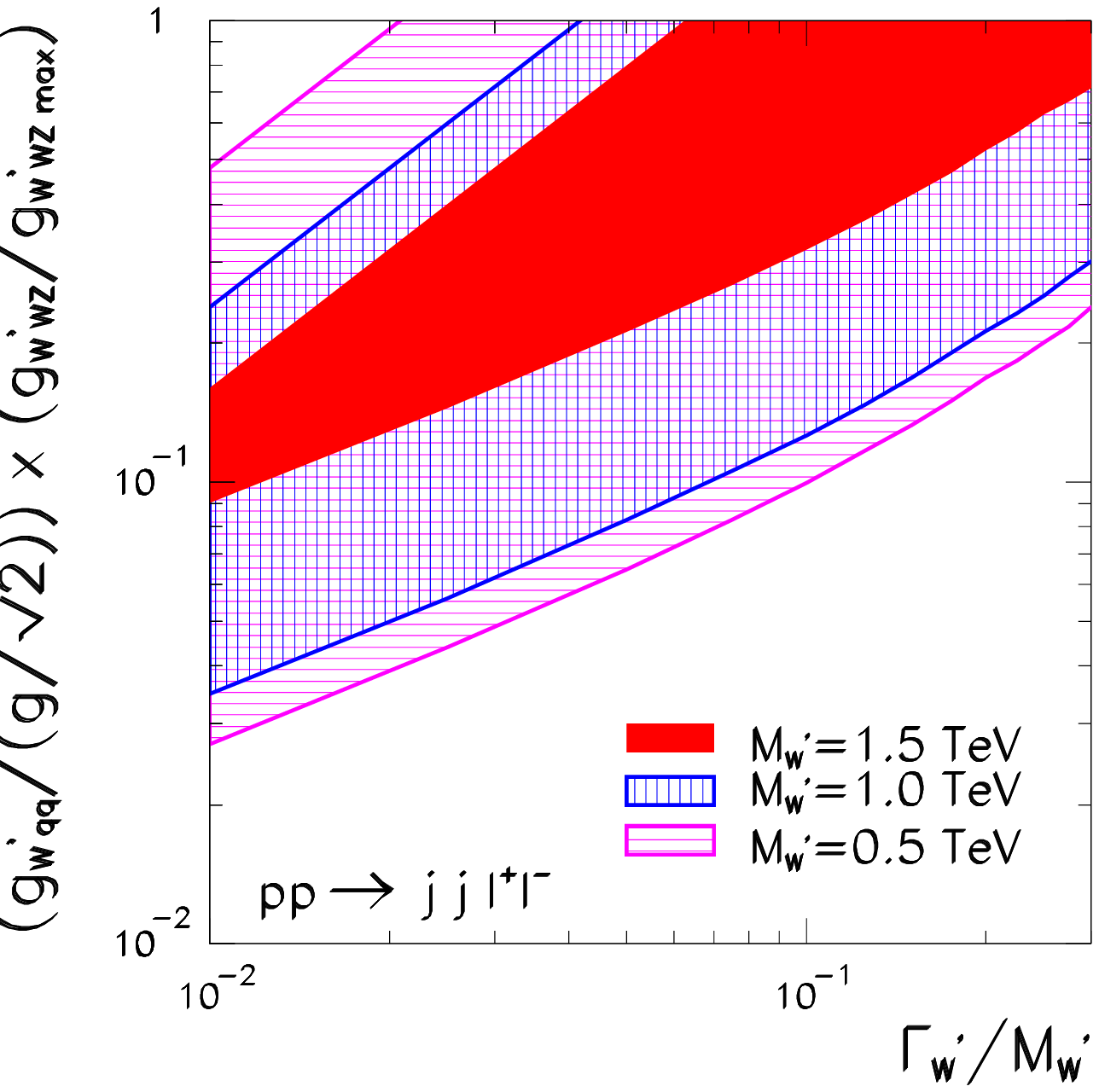}
\caption{Same as Fig.~\ref{fig:boundswplep} for the process 
$p p \to W^\prime \to W^\pm Z \to \ell^\pm \ell^\pm jj $.}
\label{fig:boundswplljj}
\end{figure}

\section{Results and conclusions}
\label{sec:conclu}

In the present study, we use as signal of the production of a new resonance 
$V^\prime$ ($=Z^\prime$ or $W^\prime$) the excess in the total number 
of events over the SM expectations after the application of the suitable
cuts described in the previous section.  

The analysis of the statistical significance of the signals described 
is simplified by the fact that in all channels the
interference between the SM and the $V'$ contributions,
$\sigma_{int}(M_{V^\prime},\Gamma_{V^\prime})$, is negligible for $\Gamma_{V^\prime}<
M_{V^\prime}/2$ after cuts.  In particular, 
for sufficiently high number of expected background events, the
$5\sigma$ sensitivity bounds on the $V^\prime$ properties can be obtained
assuming gaussian statistics as
\begin{equation} 
\left(\frac{g_{V^\prime q\bar q}}{g_{Vq\bar q}} \frac{g_{V^\prime WV}}{
{g_{V^\prime WV}}_{max}}\right)^2
\;> \;5 \,
\frac{\sqrt{\sigma_{SM}}}{\sqrt{\cal
L}~\sigma_{V^\prime}(M_{V^\prime}, \Gamma_{V^\prime})}\; .
\label{eq:lim}
\end{equation} 
Here we present our results for  an integrated
luminosity of 100 fb$^{-1}$.  For this luminosity the number of
background events is large enough to gaussian statistics to hold for
all cases with the exception of the leptonic heavy $W^\prime$ channel, $p
p \to W^\prime \to W^\pm Z \to  \ell^{\prime \pm} \ell^+ \ell^- \,
\sla{E}_T$. In this case condition (\ref{eq:lim}) is modified by
adopting the corresponding 5$\sigma$ observability bound for Poisson
statistics in the presence of the corresponding expected
background. \medskip

We depict in Figs.~\ref{fig:boundsv0lep}--\ref{fig:boundswplljj} the
ranges of couplings $g_{V^\prime q\bar q}\, g_{V^\prime WV}$ for which a 5$\sigma$
signal can be observed as a function of the resonance width for three
different values of its mass $M_{V^\prime}=0.5,~1,~1.5$ TeV in the different
channels.  There are some basic features that are common to all the
cases. First of all, the observability regions are bounded both from
below and above.  The lower bound arises from the condition of the
signal being observable while the upper limit originates from the
consistency of our effective description of the vector resonance
production. For a given vector mass the contribution of the $V^\prime \to VV$ and
$V^\prime \to q \bar{q}^\prime$ partial widths should not surpass the assumed
value of the total width; this fact is described by the upper limits
given in Eqs.~(\ref{eq:zcouplimit}) and (\ref{eq:wcouplimit}). \medskip

Another general feature is that the lower bound on the $g_{V^\prime q\bar
  q}\, g_{V^\prime WV}$ increases as the total width $\Gamma_{V^\prime}$ is augmented
just because broad resonances tend to produce more events outside the
$V^\prime$ invariant mass window used to define the signal; see for instance
(\ref{wzjjl:c4}).  Moreover, as can be expected, the lower limit on
the coupling constant product increases for heavier $V^\prime$ masses. Here,
we have two effects collaborating: not only the the production cross
section decreases for heavier vector masses, but also its decay
products are more collimated becoming harder to pass the isolation cuts
imposed on jets and leptons.  \medskip

It is interesting to compare the expected sensitivity for the $Z^\prime$ in
the purely leptonic $ \ell^+ \ell^- \,\sla{E}_T$ final state, shown in
Fig.~\ref{fig:boundsv0lep}, with the semileptonic final state,
$\ell^\pm jj \,\sla{E}_T$, presented in Fig.~\ref{fig:boundsv0nljj}.
From these figures, we see that the possibility of reconstructing the
$Z^\prime$ peak and the larger event rates leads to better sensitivity for
the semileptonic channel for $M_{Z^\prime}=0.5$ and $ 1$ TeV despite the
larger QCD backgrounds. Nevertheless, the importance of the channels
is reversed for heavier $M_{Z^\prime}$ because the $W$'s are very energetic
and their decay jets are highly collimated not passing the
isolation cut in (\ref{jjlnu:c1}).  \medskip

As for the expected sensitivity to the $W^\prime$, the best channel for all
masses considered is the purely leptonic final state $\ell^{\prime
  \pm} \ell^+ \ell^- \, \sla{E}_T$ as consequence of the achievable
background reduction. Also, as foreseeable, the predicted sensitivity to
$W^\prime$ in the $\ell^\pm jj \,\sla{E}_T$ final state is comparable to the
corresponding one for $Z^\prime$ in the same final state (up to the
difference strength in the couplings to light quarks of the $W$ and
$Z$) and only slightly better than the observability of $W^\prime$ in the
$\ell^+ \ell^- jj$ channel. \medskip

The purely leptonic and semi-leptonic channels are complementary since they 
allow us to determine which resonance ($W^\prime$ or $Z^\prime$) has been 
produced if a signal is indeed observed. The $W^\prime$ and $Z^\prime$ 
productions lead to the same final state $\ell^\pm jj \,\sla{E}_T$, however, 
additional signals in the $\ell^+ \ell^- jj$  and $\ell^{\prime \pm} \ell^+ \ell^-  \,\sla{E}_T$ channels characterize the $W^\prime$ existence, while the smoking 
gun for the $Z^\prime$ is an additional signal in the $\ell^+ \ell^{\prime -} 
\,\sla{E}_T$ channel. \medskip

We also notice that should the narrow width approximation hold for
$\sigma_{V^\prime}$ one would have
\begin{equation} \sigma_{V'}\propto \sigma(pp\to V^\prime) \times
Br(V^\prime \rightarrow WV) = A \frac{(g_{V^\prime q\bar q}\,
g_{V^\prime WV})^2}{\Gamma_{V^\prime}}
\label{eq:bandwidth}
\end{equation} where $A$ depends only on $M_{V^\prime}$.  Thus, in the region
of validity of the narrow width approximation, the bounds on $V^\prime$ obey
a simple scaling law characterized by
\begin{equation} \left( \frac{g_{V^\prime q\bar q}}{g_{Vq\bar q}} ~
\frac{g_{V^\prime WV}}{g_{V^\prime WV~max}} 
\right )^2\, \times \Gamma_{V^\prime}
\label{eq:boundwidth2}
\end{equation} being a constant depending on $M_{V^\prime}$ and the final
state considered.  As we can see in Figs.~\ref{fig:boundsv0lep} to
\ref{fig:boundswplljj} our results are reasonably well described by
Eq.~(\ref{eq:boundwidth2}) for $\Gamma_{Z^\prime}\lesssim 0.05$ --
$0.1M_{Z^\prime}$, while there is a departure from this simple scaling rule
for larger widths. \medskip

As a final comment, let's notice that these results should be taken 
with a pitch of salt since there might
exist additional backgrounds to the vector resonance searches coming
from the new physics associated to the electroweak symmetry breaking
sector.  For instance, in extra dimension models there is a plethora
of new heavy states that might give rise to final states similar to
the ones we studied, {\em e.g.}  KK charged leptons can decay into
a $W$ pair accompanied by neutrinos. If these backgrounds turned out
to be sizeable, certainly the reconstruction of the new resonance
Breit--Wigner profile would become an important tool to isolate the
new vector states. Moreover, we have not considered the 
systematic uncertainties
associated to the prediction of the absolute value of the SM 
backgrounds that can be of the order of 5--20\%~\cite{Campbell:2009kg}.

\medskip
 
In summary, in this work we have performed a model independent analyses 
of the observability  of a new spin--1 particle, either neutral or
charged, that is associated to the unitarity restoration in the
electroweak vector boson scattering. In addition to the 
essential couplings to electroweak gauge boson pairs we also
considered the new vector resonance couplings to light
quarks; many extensions of the SM, like higgsless models, contain
such couplings. We analyzed the five  channels shown in Eqs.~(\ref{ppww})
and (\ref{ppwz}) expressing the attainable bounds as a function
of the resonance effective couplings, its mass and width. Our analyses show 
that the study of these process have a large reach at the LHC, allowing to
discover or exclude many scenarios like Higgsless models. \medskip

\section*{Acknowledgments}
M.C. G-G thanks the CERN Theory Group for their hospitality
during the final stages of this work. 
This work was partially supported by Conselho Nacional de
Desenvolvimento Cient\'{\i}fico e Tecnol\'ogico (CNPq) and by
Funda\c{c}\~ao de Amparo \`a Pesquisa do Estado de S\~ao Paulo
(FAPESP).  M.C. G-G is supported by the National Science Foundation
under Grant PHY-0354776 and by Spanish Grants FPA-2007-66665-C02-01
and CSD2008-0037.


\bibliographystyle{h-physrev4}

\end{document}